\documentclass[prd,aps,preprint,tightenlines,showpacs,nofootinbib,superscriptaddress]{revtex4-1}
\usepackage{mathrsfs}
\usepackage{amsfonts}
\usepackage{amsmath}
\usepackage{amssymb}
\usepackage{array}
\usepackage{verbatim}
\usepackage{bm}
\usepackage{epsfig}
\usepackage{graphicx,color}
\usepackage{relsize}
\usepackage{lineno}
\usepackage{float}
\usepackage{multirow}
\RequirePackage{xspace}

\def\pom{{I\!\!P}}

\begin{document}

\begin{flushright}
MS-TP-23-25
\end{flushright}

\title{\boldmath Exclusive $J/\Psi$ plus jet associated production \\ in ultraperipheral $PbPb$ collisions}

\author{Victor P. Gon\c{c}alves}
\email{barros@ufpel.edu.br}
\affiliation{Institut f\"ur Theoretische Physik, Westf\"alische Wilhelms-Universit\"at M\"unster,
Wilhelm-Klemm-Straße 9, D-48149 M\"unster, Germany
}
\affiliation{Physics and Mathematics Institute, Federal University of Pelotas, \\
  Postal Code 354,  96010-900, Pelotas, RS, Brazil}

\author{Michael { Klasen}}
\email{michael.klasen@uni-muenster.de}
\affiliation{Institut f\"ur Theoretische Physik, Westf\"alische Wilhelms-Universit\"at M\"unster,
Wilhelm-Klemm-Straße 9, D-48149 M\"unster, Germany
}

\author{Bruno D. { Moreira}}
\email{bduartesm@gmail.com}
\affiliation{Departamento de F\'isica, Universidade do Estado de Santa Catarina, 89219-710 Joinville, SC, Brazil.}

\begin{abstract}
The study of exclusive processes in  ultraperipheral collisions at the Large Hadron Collider (LHC) has allowed us to test several aspects of the Standard Model and to search for New Physics. In this letter, we investigate the possibility of using these processes to improve our understanding of the  quarkonium production mechanism through the study of the exclusive $J/\Psi$ plus jet associate production in ultraperipheral $PbPb$ collisions. We estimate the transverse -  momentum and rapidity distributions considering that the $\gamma \gamma \rightarrow J/\Psi + X$ ($X = \gamma, \, g$) subprocess is described by the Non - Relativistic QCD (NRQCD) formalism and present predictions for the rapidity ranges covered by central and forward detectors. The experimental separation of these events is discussed and the results indicate that a future experimental analysis is, in principle, feasible in future runs of the LHC and the Future Circular Collider (FCC).
\end{abstract}

\maketitle

\section{Introduction}
  
Over the last decades, the study of photon – induced interactions in hadronic colliders became a reality, such that
currently the Large Hadron Collider (LHC) is also considered a powerful photon – hadron and photon – photon collider, which can be used to improve our understanding of the Standard Model as well as to searching for New Physics (for reviews see, e.g. Refs. \cite{Budnev:1975poe,upc,upc2}). One of the more promising processes is the exclusive particle production in ultraperipheral heavy ion collisions (UPHICs), which are characterized by an impact parameter that is larger than the sum of the radii of the incident particles. In exclusive processes the final state is very clean, being characterized by the produced system, two intact ions and two rapidity gaps, i.e. the outgoing particles are separated by a large region of rapidity in which there is no additional hadronic activity observed in the detector. 
In particular,  the exclusive quarkonium photoproduction in photon - Pomeron ($\gamma \pom$) interactions has been largely studied, mainly motivated by the possibility to constrain the description of the QCD dynamics at high energies and of quarkonium production, as well as to improve our understanding of the quantum 3D imaging of the partons inside the protons and nuclei \cite{upc2}. In this case, the nucleus scatters elastically and 
remains intact in the final state, with the quarkonium being produced with a small squared transverse momentum
($p_t^2 \lessapprox 1/R_A^{2}$, where $R_A$ is the nuclear radius) due to the coherence condition, and with its   signature being a sharp forward diffraction peak in the transverse - momentum distribution, which has a typical $e^{-cte.p_t^2}$ behaviour. The current data have already  constrained several aspects of the theoretical description of this process and 	 more precise measurements are expected in the forthcoming years, especially at larger transverse momentum, where incoherent processes, in which the nucleus scatters inelastically and breaks-up, are expected to dominate.

\begin{figure}
	\centering
	\includegraphics[width=5in]{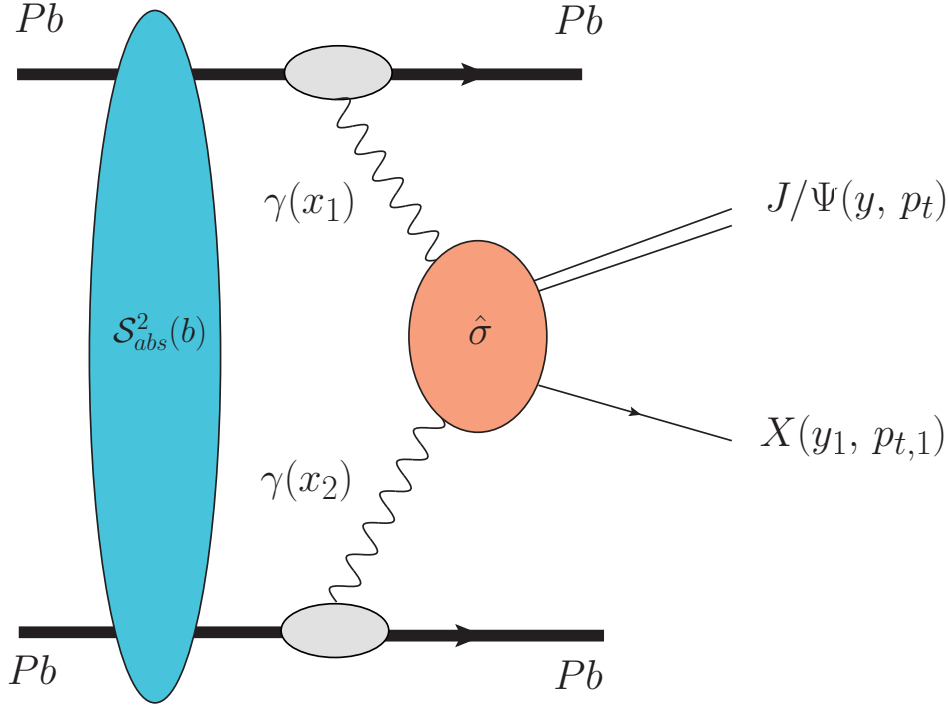}
	\caption{Exclusive $J/\Psi + X$ ($X = \gamma,\,g$) production in ultraperipheral $PbPb$ collisions.		 }
	\label{Fig:diagram}
\end{figure}

In this letter we will investigate another possibility for the exclusive quarkonium production in UPHICs, which has not been previously analyzed in detail in the literature: the exclusive quarkonium production by photon - photon ($\gamma \gamma$) interactions, represented in Fig. \ref{Fig:diagram} for the $J/\Psi$ case. In this process, the quarkonium is produced in association with a hard photon or a gluon jet, with the transverse momentum distribution of the meson having a power - like behaviour ($\propto 1/p_t^n$), characteristic of a $2 \rightarrow 2$ process. One has that in UPHICs the $\gamma \gamma$ interactions are  enhanced by a factor $Z^4$ \cite{upc}, which implies that the resulting cross sections are expected to be non - negligible, and its experimental signature will be identical to that generated in $\gamma \pom$ interactions if the associated system $X = \gamma, \, g$ is produced in a rapidity range not covered by the LHC detectors. Moreover, this process is expected to dominate if a minimum value for the transverse momentum of the meson is imposed in the experimental analysis. Our goal in this letter is twofold. First, to estimate the transverse - momentum and rapidity distributions for the $PbPb \rightarrow Pb \otimes J/\Psi X \otimes Pb$ process, where $\otimes$ indicates the presence of a rapidity gap in the final state, without any restriction on the rapidity of the associated system $X$. Predictions for $X = \gamma$ or $g$ will be presented, as well as for the sum of these two contributions. 
Second, and more important, to perform a detailed analysis of the cross section for the case where the system $X$ is imposed to have a rapidity $y_1$ in the kinematical range covered by central ($-2.5 \le y_1 \le +2.5$) and forward ($+2.0 \le y_1 \le +4.5$) detectors. The presence of a hard photon or gluon jet in association to the $J/\Psi$ meson in the detector implies that the background due to $\gamma \pom$ interactions is strongly reduced. As a consequence, in this situation, the analysis of the  $PbPb \rightarrow Pb \otimes J/\Psi X \otimes Pb$ process can be used to study the quarkonium production mechanism, which is still one of the main challenges in Particle Physics (for a recent review see, e.g., Ref. \cite{Lansberg:2019adr}).

 This letter is organized as follows. In the next Section, we will present a brief review of the formalism needed to describe the exclusive $J/\Psi + X$ production by $\gamma \gamma$ interactions in $PbPb$ collisions. In Section \ref{sec:res} we will present our predictions for the differential distributions considering the associated production of the $J/\Psi$ with a hard photon or a gluon, as well as for the sum of these two contributions. Moreover, predictions for the transverse momentum and rapidity distributions will be presented for the case in which the associated system $X$ has a rapidity in the kinematical range covered by central and forward detectors. Results for the total cross sections considering $PbPb$ collisions for the energies of the Large Hadron Collider (LHC) and Future Circular Collider (FCC) will also be presented. Finally, in Section \ref{sec:sum}, we will summarize our main results and conclusions.

\section{Formalism}

The treatment of quarkonium production has attracted much attention during the last decades, mainly motivated by the possibility to probe the interplay between the long and short distance regimes of the strong interactions (for a recent review see, e.g., Ref. \cite{Lansberg:2019adr}).
 One of the more successful approaches is the Non - Relativistic QCD (NRQCD) formalism \cite{Bodwin:1994jh}, in which  the cross section for the production of a heavy quarkonium state $H$ factorizes as  $\sigma (ab \rightarrow H+X)=\sum_n \sigma(ab \rightarrow Q\bar{Q}[n] + X) \langle {\cal{O}}^H[n]\rangle$, where the coefficients $\sigma(ab \rightarrow Q\bar{Q}[n] + X)$ are perturbatively calculated short - distance cross sections for the production of the heavy quark pair $Q\bar{Q}$ in an intermediate Fock state $n$, which does not have to be color neutral.  The $\langle {\cal{O}}^H[n]\rangle$
are nonperturbative long distance matrix elements (LDMEs), which describe the transition of the intermediate $Q\bar{Q}$ in the physical state $H$ via soft gluon radiation. Currently, these elements have to be extracted in a global fit to quarkonium data, as performed, for instance, in Refs. \cite{Butenschoen:2011yh,Chao:2012iv,Gong:2012ug}. 

The NRQCD formalism was applied for the $J/\Psi + X$ production in two - photon collisions  in Refs. \cite{Klasen:2001mi,Klasen:2004tz,Klasen:2004az}, which have analyzed in detail the production of this final state in $e^+ e^-$ collisions. Such studies have demonstrated that the experimental analysis of this process would be useful to verify the NRQCD factorization for charmonium, to provide the existence of Color Octet processes and to constrain the associated LDMEs. 
Unfortunately, $e^+ e^-$ colliders are not yet a reality, and data in these collisions are only expected to be available more than two decades ahead. An alternative, which we analyze in this letter, is to study this process in the LHC considering ultraperipheral heavy ion collisions \cite{upc}. In these collisions, the two charged nuclei interact at impact
parameters larger than the sum of their radii, with the incident ions acting as sources of almost real photons, which implies that photon -- photon interactions may happen. Although the center - of - mass energies for $\gamma \gamma$ interactions achieved in $PbPb$ collisions are smaller than those expected in future $e^+ e^-$ collisions, in heavy ion collisions the cross sections are enhanced by a factor $Z^4$ associated to the $Z^2$ dependence of the photon flux. 
In what follows, we will present a brief review of the formalism.

Assuming the validity of the equivalent photon approximation (EPA)\cite{Budnev:1975poe}, one has that the differential cross section for the $PbPb \rightarrow Pb \otimes J/\Psi\,X \otimes Pb$ process, represented in Fig. \ref{Fig:diagram}, can be expressed as follows
\begin{eqnarray}
\frac{d\sigma}{d^2p_t dy dy_1} = \int d^2{\mathbf r_1} d^2{\mathbf r_2} \, x_1 f^{\gamma}_{Pb}(x_1,{\mathbf r_{1}}) \, x_2 f^{\gamma}_{Pb}(x_2,{\mathbf r_{2}}) \sum_n \sum_{a = \gamma,\,g} \frac{d \hat{\sigma}}{dt} [\gamma \gamma \rightarrow c\bar{c}(n)\,a;\,W]S^2_{abs}(\mathbf b) \,\,,
\label{Eq:cs}
\end{eqnarray}
where $p_t$ and $y$ are the transverse momentum and rapidity of the $J/\Psi$ meson and $y_1$ the rapidity of the gluon jet or hard photon. Moreover,  $x f^{\gamma}_A(x,{\mathbf r})$ is the photon flux produced by a relativistic charge $Z$ at transverse distance $r \equiv |{\mathbf r}|$ from its center, which is given by \cite{Budnev:1975poe,upc}
\begin{eqnarray}
x f^{\gamma}_A(x,{\mathbf r})  = \frac{Z^{2}\alpha_{em}}{\pi^2}\frac{1 + (1-x)^2}{2}
\cdot \left|
\int \frac{dk_{\perp} k_{\perp}^{2}}{k_{\perp}^2 + (x \, m_p)^2} J_{1}(r k_{\perp}) 
F\left( k_{\perp}^2 + (x \, m_p)^2\right)\right|^{2} \,\,,
\label{fluxo}
\end{eqnarray}
 where $x$ is the fraction of the ion energy carried by the photon, $k_{\perp}$ is the photon tranverse momentum, $m_p$ is the proton mass and $F$ is the charged form factor of the nucleus.
In addition,  $W = \sqrt{ x_1 x_2 {s_{NN}}}$ is the  $\gamma \gamma$ center - of - mass energy,
with $x_{1} = [m_T\exp(+y) + p_{t,1}\exp(+y_1)]/\sqrt{s_{NN}}$, $x_{2} = [m_T\exp(-y) + p_{t,1}\exp(-y_1)]/\sqrt{s_{NN}}$, $m_T = \sqrt{p_t^2 + M^2}$, $p_t = p_{t,1}$ and $M$ is the $J/\Psi$ mass. The  
 differential cross section for the $\gamma \gamma \rightarrow c\bar{c}(n)\,a$ subprocess,   $d \hat{\sigma}/dt$,   is described in detail in Ref. \cite{Klasen:2001mi} and depends on the long - distance matrix elements (LDME) $\langle {\cal{O}}_1^{J/\Psi}(^3S_1)\rangle$ and $\langle {\cal{O}}_8^{J/\Psi}(^3S_1)\rangle$ for $a = \gamma$ and $a=g$, respectively. The factor $S^2_{abs}({\mathbf b})$ depends on the impact parameter ${\mathbf b} = {\mathbf r_{1}} - {\mathbf r_{2}}$ of the heavy ion collision and  is denoted the absorptive  factor, which excludes the overlap between the colliding ions and allows to take into account only ultraperipheral collisions, characterized by $|{\mathbf b}| \ge 2 R_{Pb}$, where $R_{Pb}$ is the lead ion radius.

In our analysis, we will estimate the associated photon spectrum considering a realistic form factor, which corresponds to the Wood - Saxon distribution and is the Fourier transform of the charge density of the nucleus, constrained by the experimental data \cite{DeJager:1974liz}. Moreover, the absorptive factor will  be estimated using the Glauber approach discussed in Refs. \cite{klein,Baltz:2009jk}. For the LDME we will use the values obtained in Ref. \cite{Butenschoen:2011yh}: $\langle {\cal{O}}_1^{J/\Psi}(^3S_1)\rangle = 1.32$ GeV$^3$ and $\langle {\cal{O}}_8^{J/\Psi}(^3S_1)\rangle = 0.0022$ GeV$^3$.

\begin{figure}
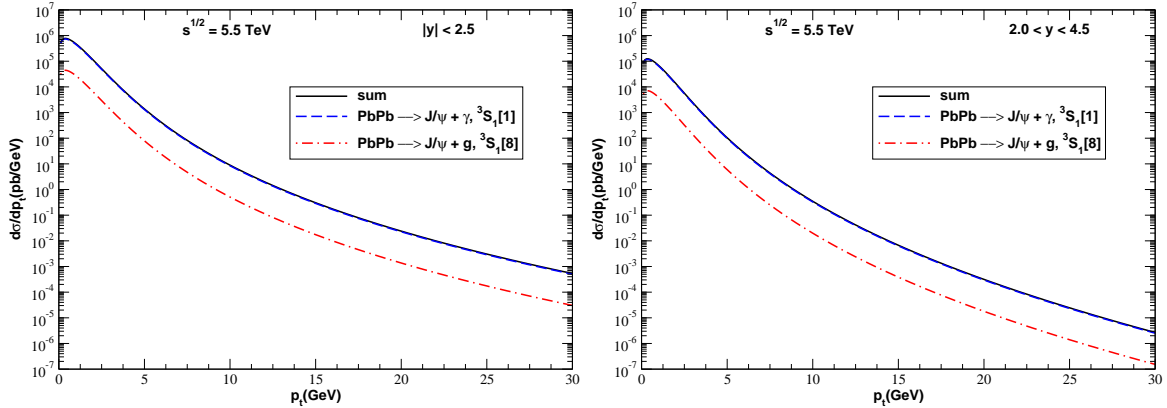

	\centering
	\includegraphics[width=3in]{dist_pt_psi_PbPb_LHC_central.eps}
	\includegraphics[width=3in]{dist_pt_psi_PbPb_LHC_forward.eps}
		\caption{Transverse momentum distributions for the exclusive $J/\Psi + X$ production in $PbPb$ collisions at the LHC ($\sqrt{s_{NN}} = 5.5$ TeV) considering that the $J/\Psi$ meson is produced in the rapidity range covered by a central (left panel) or a forward (right panel) detector. No restriction is imposed in the rapidity of the system $X$.}
	\label{Fig:dsdptfully1}
\end{figure}

\section{Results}
\label{sec:res}
In what follows we will present our predictions for the differential cross sections associated with the $PbPb \rightarrow Pb \otimes J/\Psi + X \otimes Pb$ process at the LHC energy ($\sqrt{s_{NN}} = 5.5$ TeV) and considering that $X = \gamma$, a gluon jet $g$ or the sum of both contributions. As discussed in Ref. \cite{Klasen:2001mi}, the accessible phase space is defined by
\begin{eqnarray}
& \,& 0 \le  p_t  \le \frac{s_{NN} - M^2}{2\sqrt{s_{NN}}} \,,\\
& \,& |y|  \le  {\mbox{Arccosh}} \left( \frac{s_{NN}+M^2}{2\sqrt{s_{NN}}m_T} \right) \,,\\
& \,& - \ln \frac{\sqrt{s_{NN}}-m_T\exp(-y)}{p_{t,1}} \le  y_1 \le + \ln \frac{\sqrt{s_{NN}}-m_T\exp(y)}{p_{t,1}} \,\,.
\end{eqnarray} 
Initially, let's assume that the meson is produced in the rapidity range covered by central ($-2.5 \le y \le +2.5$) and forward ($+2.0 \le y \le +4.5$) detectors and no restriction will be imposed on the rapidity $y_1$ of the  system $X$.   
The resulting predictions are presented in Fig. \ref{Fig:dsdptfully1}. One has that the shape of the distributions for $X = \gamma$ and $g$ are very similar, which is expected since the production of these two final states receives contributions from the same Feynman diagrams. However, the normalization is different due to the distinct couplings and LDMEs. The process is dominated by the photon production, which is associated with the fact that one has a color singlet configuration. In contrast, the associated production with a gluon only is possible if one takes into account the color octet contribution, since the direct production of a color - singlet with one gluon is forbidden by color  conservation.  One has that the maximum of the distribution occurs for $p_t \approx 1.0$ GeV and it has a power - like behavior, decreasing for larger values of the transverse momentum, as expected for a $2 \rightarrow 2$ process. This decrease is faster for forward rapidities, as well the normalization is smaller in comparison to central rapidities.

\begin{figure}
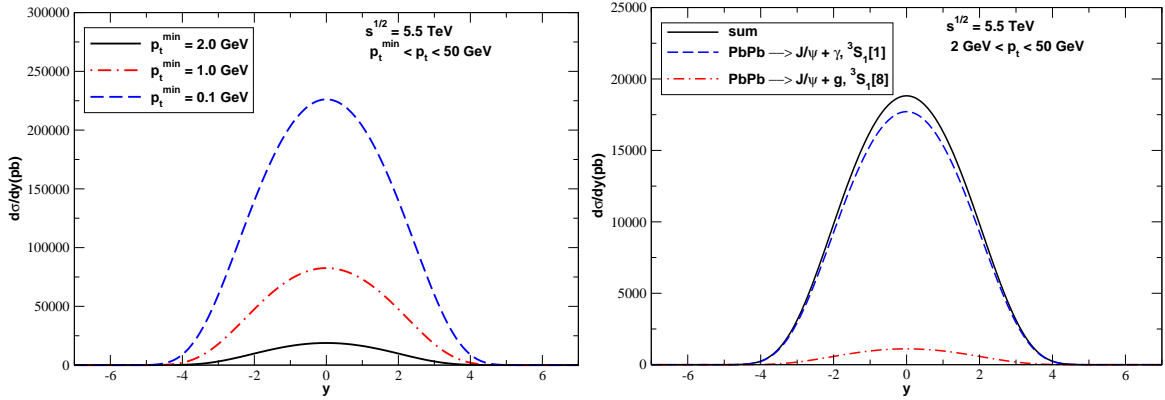

	\centering
	\includegraphics[width=3in]{dist_y_psi_LHC_ptmin.eps}
	\includegraphics[width=3in]{dist_y_psi_LHCv2.eps}
	\caption{{\bf Left panel:} Dependence on the minimum transverse momentum $p_t^{min}$ of the rapidity distribution for the exclusive $J/\Psi + X$  production in ultraperipheral $PbPb$ collisions at the LHC energy.		{\bf Right panel:} Predictions for the rapidity distribution considering $X = \gamma$ and $X = g$ derived assuming $p_{t}^{min} = 2.0$ GeV. The sum of the contributions is also presented for comparison. }
	\label{Fig:ptmin}
\end{figure}

In order to estimate the rapidity distribution one must integrate Eq. (\ref{Eq:cs}) over $y_1$ and over  the meson transverse momentum $p_t$. Theoretically, we can assume   a very small value for the lower integration limit since the distribution is well - behaved for $p_t \rightarrow 0$. However, experimentally, this value is determined by the capability of the detector to measure particles with small values of $p_t$. The results from ALICE and LHCb Collaborations have demonstrated that they are able to select exclusive events characterized by a meson with very small transverse momentum. However, as the background associated to the exclusive $J/\Psi$ production by $\gamma \pom$ interactions is strongly reduced by selecting events with $p_t \gtrsim 0.1$ GeV, we present in Fig. \ref{Fig:ptmin} (left panel) the predictions derived considering three different values of $p_{t}^{min}$ and summing the contributions for $X = \gamma$ and $X = g$. 
One has that the normalization of the distribution for $y = 0$ is strongly dependent on $p_{t}^{min}$. In particular, the distribution decreases by one order of magnitude when it is increased from 0.1 to 2.0 GeV.
In what follows, we will assume $p_{t}^{min} = 2.0$ GeV in our calculations, since we are interested in events where the associated photon or gluon jet can also be measured. However, it is important to keep in mind that our predictions can be enhanced if a smaller value can be accessed by the experimental groups.
In Fig. \ref{Fig:ptmin} (right panel) we present the predictions for the rapidity distribution for the exclusive $J/\Psi + X$ production considering $X = \gamma$ and $X = g$.  The sum of the contributions is also presented for comparison. As expected from Fig. \ref{Fig:dsdptfully1}, the $J/\Psi + \gamma$ process dominates, with its normalization for $y = 0$ being a factor $\approx 8$ larger than for the  $J/\Psi + g$ production.

\begin{table}
    \centering
    \begin{tabular}{|l|l|l|l|}
    \hline
        ~ & full $y,y_{1}$ & $|y|<2.5,y_{1}$ full &  $2.0<y<4.5,y_{1}$ full  \\ \hline
        {\bf LHC ($\sqrt{s_{NN}} = 5.5$ TeV)} & 77.87 nb & 70.53 nb &   7.67 nb \\ \hline
        {\bf FCC ($\sqrt{s_{NN}} = 39.0$ TeV)} & 384.77 nb & 268.10 nb &  74.83 nb  \\ \hline
    \end{tabular}
    \caption{Predictions for the total cross sections considering the exclusive $J/\Psi + X$ production in $PbPb$ collisions at the LHC and FCC energies and derived assuming different  ranges for the meson rapidity. The rapidity of  $X$ has been integrated over the full rapidity range.}
    \label{Tab:fully1}
\end{table}

In Table \ref{Tab:fully1} we present the cross sections for the exclusive $J/\Psi + X$ production in $PbPb$ collisions at the LHC and FCC energies,  derived assuming different  ranges for the meson rapidity and integrating  $y_1$ over the full rapidity range.
For comparison the results derived when no restriction on $y$ is assumed are also presented. One has that the cross section increases with the center - of - mass energy, being of the order of dozens (hundreds) of nb for the LHC (FCC) energy for central rapidities. For the LHC energy, the cross section decreases by $\approx 10$ \% when a central rapidity cut is assumed and by one order of magnitude if the meson is assumed to be produced at forward rapidity. In contrast, for the FCC energy\footnote{The predictions for the rapidity and transverse momentum distributions derived considering $PbPb$ collisions at the FCC energy are available upon request. }, the reduction is of $\approx 30$ \%, which is expected since the distribution becomes wider with the increasing of $\sqrt{s_{NN}}$. Considering the high luminosities expected in the future runs of LHC and FCC, these results imply a non-negligible number of events, which can be enriched if a smaller value of $p_{t}^{min}$ is assumed. 

The previous results indicate that the contribution of the $J/\Psi + X$ process for the exclusive production of a $J/\Psi$ meson with $p_t \ge p_{t}^{min}$ can be important at the LHC. In principle, if the hard photon or gluon jet are produced in the rapidity range covered by the ALICE and LHCb detectors,  the contribution of this process can be eliminated by imposing a cut on the number of additional tracks on detector, as usually made in the experimental analysis of the exclusive $J/\Psi$ production. If the system $X$ is produced in a region not covered by the detectors, the $J/\Psi + X$ process is a background for the events observed at large $p_t$. On the other hand, if we impose as a selection  criterion the measurement of a meson and a hard photon or gluon jet, with $\vec{p}_t \approx -  \vec{p}_{t,1}$, we can suppress the contribution of the $\gamma \pom$ interactions and have direct access to the $J/\Psi + X$ process, which would allow us to study the quarkonium production mechanism. In what follows, we will derive predictions for the case when this selection criterion is assumed.

\begin{figure}
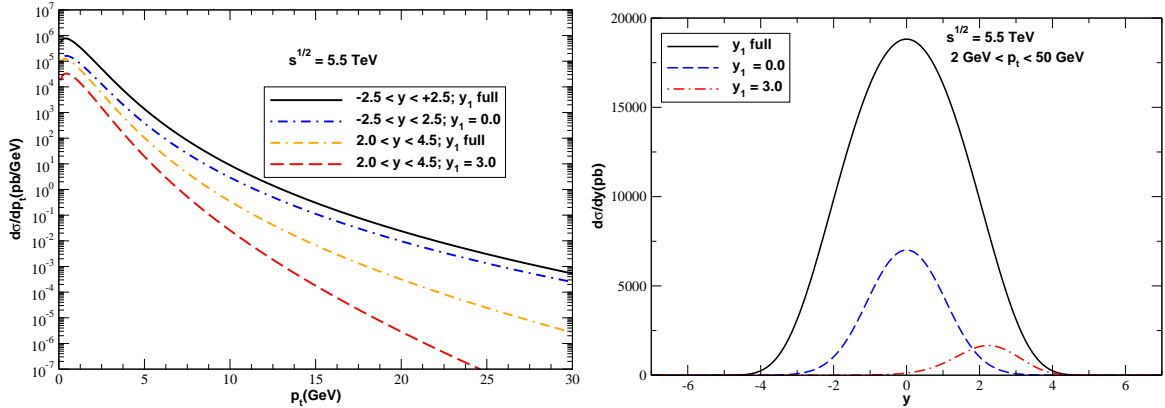

	\centering
	\includegraphics[width=3in]{dist_pt_psi_PbPb_LHC_fixedy1.eps}
	\includegraphics[width=3in]{dist_y_psi_LHC_fixedy1.eps}
	\caption{{\bf Left panel:} Predictions for the transverse momentum distributions derived considering different ranges for the meson rapidity and distinct values for the rapidity of the system $X$.	{\bf Right panel:} Predictions for the rapidity distribution considering distinct values for the rapidity of system $X$. The result for the case where $y_1$ is integrated over the accessible phase space is also presented for comparison.}
	\label{Fig:fixedy1}
\end{figure}

In Fig. \ref{Fig:fixedy1} (left panel) we present our predictions for the transverse momentum distribution derived considering different ranges for the meson rapidity $y$ and distinct values for the rapidity of system $X$. For comparison, the results derived when $y_1$ is integrated over the accessible phase space are also shown.
For a $J/\Psi$ meson in the central detector, one has that imposing the presence of a hard photon or a gluon jet at midrapidity decreases the normalization of the distribution, but its shape is not modified. On the other hand, if we impose that the $J/\Psi$ meson is detected at forward rapidity, the normalization and the shape of the distribution are modified by imposing that the system $X$ is also present at forward rapidity. Such a result is associated with the  reduction of the accessible phase space when we increase the rapidities of the final states. 

The predictions for the rapidity distribution, derived for different values of $y_1$, are presented in Fig. \ref{Fig:fixedy1} (right panel). The result for the case where $y_1$ is integrated over the accessible phase space is also presented for comparison. One has that imposing the presence of the system $X$ at $y_1 = 0$ reduces the normalization of the distribution for midrapidity by a factor of $\approx 3$, but its shape remains symmetric over $y = 0$, with the maximum occurring in the kinematical range covered by the ALICE detector. In contrast, for the case where we assume $y_1 = 3$, the normalization and the position of the maximum of $d\sigma/dy$ are modified, with the peak occuring for $y \approx 2.2$, i.e. in the kinematical range covered by the LHCb detector.

\begin{table}[t]
    \centering
    \begin{tabular}{|l|l|l|l|}
    \hline
        ~ & full $y,y_{1}$ &  $|y|<2.5,$ $y_{1}=0$ &  $2.0<y<4.5,y_{1}=3$ \\ \hline
        {\bf LHC ($\sqrt{s_{NN}} = 5.5$ TeV)} & 77.87 nb  & 17.97 nb &   2.07 nb\\ \hline
        {\bf FCC ($\sqrt{s_{NN}} = 39.0$ TeV)} & 384.77 nb &  57.33 nb &  23.85 nb  \\ \hline
    \end{tabular}
    \caption{Predictions for the total cross sections considering the exclusive $J/\Psi + X$ production in $PbPb$ collisions at the LHC and FCC energies and derived assuming different  ranges for the meson rapidity and fixed values for the rapidity of the system $X$.}
    \label{Tab:fixedy1}
\end{table}

Finally, in Table \ref{Tab:fixedy1} we present the predictions for the total cross sections associated to the exclusive $J/\Psi + X$ production in $PbPb$ collisions at the LHC and FCC energies,  derived assuming different  ranges for the meson rapidity and fixed values for the rapidity of the system $X$.
In comparison to the case where the rapidities of the final states are integrated over the accessible phase space, one has that if we impose that both the $J/\Psi$ and $X$ are measured by a central detector, the predictions for LHC energy are reduced by a factor of $\approx 4$. Such a factor is $\approx 35$ for a forward detector. At the FCC energy, we predict larger values of the cross sections by a factor of $\approx 3$ (10) for a central (forward) detector. Such results indicate that a future experimental analysis of the exclusive
 $J/\Psi + X$ production in $PbPb$ collisions is, in principle, feasible.

\section{Summary}
\label{sec:sum}
The description of the quarkonium production mechanism is still one of the main challenges  of Particle Physics. In recent years, our understanding about the subject has largely improved, mainly associated with the large amount of data released by the LHC Collaborations for quark and gluon initiated processes. Currently, there is the expectation that the future $ep$ and $e^+e^-$ colliders will allow us to convincingly establish the NRQCD formalism through the measurement of the singlet and color octet contributions expected to be present in photon - parton and photon - photon interactions.  In this exploratory study, we have investigated the alternative of studying the quarkonium production mechanism using the exclusive $J/\Psi$ plus jet associated production in ultraperipheral $PbPb$ collisions at the LHC and FCC. Our analysis was motivated by the possibility of separating these events by imposing a cut on the  meson transverse momentum and/or by measuring the associated hard photon or gluon jet. We have derived predictions for the total cross sections, transverse momentum and rapidity distributions considering the rapidity kinematical ranges covered by central and forward detectors. Our results indicate that a future experimental analysis is, in principle, feasible, which strongly motivates a more detailed analysis taking into account the  cuts usually considered by the LHC Collaborations.

\section*{Acknowledgments}
 V.P.G. was partially supported by  CNPq, CAPES, FAPERGS and  INCT-FNA (process number 
464898/2014-5). The research of M.K. was supported by the DFG through the Research
Training Group 2149 "Strong and Weak Interactions - from Hadrons to Dark
Matter" and the SFB 1225 "Isoquant", project-id 273811115, and by the BMBF
under contract 05P21PMCAA.

\end{document}